\documentclass[prb,twocolumn,showpacs]{revtex4-1}

\usepackage{graphicx}
\usepackage{multirow}
\usepackage{dcolumn}
\usepackage{bm}
\usepackage[normalem]{ulem}
\usepackage{amsmath}
\usepackage{amsfonts}
\usepackage{amssymb}
\usepackage{color}
\usepackage{colordvi}
\usepackage{dcolumn}

\usepackage[colorlinks=true, breaklinks=true, linkcolor=blue,
urlcolor=blue, citecolor=blue]{hyperref}

\newcommand{\sgn}{\mathrm{sgn}}

\begin{document}

\title{Edge effects in the magnetic interference pattern of a ballistic SNS junction}
\date{\today }
\author{Hendrik Meier,$^1$ Vladimir I. Fal'ko,$^2$ and Leonid I. Glazman$^1$}
\affiliation{
$^1$Department of Physics, Yale University, New Haven, Connecticut 06520, USA\\
$^2$National Graphene Institute, University of Manchester, Manchester, M13 9PL, United Kingdom
}

\begin{abstract}
We investigate the Josephson critical current $I_c(\Phi)$ of a wide superconductor-normal metal-superconductor (SNS) junction as a function of the magnetic flux~$\Phi$ threading it. Electronic trajectories reflected from the side edges alter the function~$I_c(\Phi)$ as compared to the conventional Fraunhofer-type dependence. At weak magnetic fields, $B\lesssim \Phi_0/d^2$, the edge effect lifts zeros in $I_c(\Phi)$ and gradually shifts the minima of that function toward half-integer multiples of the flux quantum. At $B>\Phi_0/d^2$, the edge effect leads to an accelerated decay of the critical current~$I_c(\Phi)$ with increasing $\Phi$. At larger fields, eventually, the system is expected to cross into a regime of ``classical'' mesoscopic fluctuations that is specific for wide ballistic SNS junctions with rough edges.
\end{abstract}

\maketitle

\section{Introduction}

Since the discoveries of the Josephson effect\cite{josephson64,andersonrowell} and Andreev reflection\cite{andreev}, 
proximity structures involving one or more superconducting layers have been the subject
of intense experimental and theoretical research, providing a rich playground to 
manifestations of quantum coherence. In the condition of zero voltage bias,
direct current transport between two superconductors coupled by a junction depends 
on the phase difference~$\chi$ of the two superconductors' order parameters and 
the external magnetic flux~$\Phi$ squeezed into the space between the two superconducting leads.
The functional form of the Josephson current~$I(\chi,\Phi)$ reflects
the geometry of the junction as well as the physical properties of
the interface material\cite{bergeret_rmp,kouwenhoven,halperinyacoby15,flensberg15,yacoby14,beenakker15} and that of the superconductors.\cite{vanharlingen}
Thus, in particular 
due to the arrival of the new class of ballistic superconductor-normal metal-superconductor (SNS) systems based on encapsulated graphene,\cite{falkogeim,vandersypen} the Josephson current~$I(\chi,\Phi)$ 
is an interesting object of study.

\begin{figure}[b]
\centerline{\includegraphics[width=0.5\linewidth]{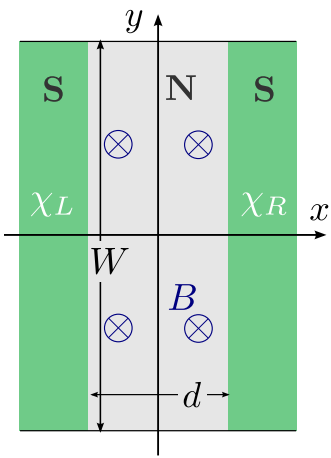}}
\caption{(Color online)
The SNS junction in an external magnetic field~$B$ under consideration. In this work, we study two-dimensional long and wide junctions, meaning that
$W\gg d$ while $d$ is larger than the microscopic lengths  
characterizing the superconducting
and normal layers.
}
\label{fig_sns}
\end{figure}

In this work, we develop a theory for the magnetic field dependence of the Josephson current in a long and wide two-dimensional ballistic 
SNS junction, see Fig.~\ref{fig_sns}. Our theory extends beyond the standard Fraunhofer interference pattern\cite{barone,tinkham} and is applicable over a broad range of magnetic fields~$B$. Overall, we find for the magnetic field dependence of the Josephson critical current the form   
\begin{align}
I_c(B) = \frac{I_{c0}}{\varphi}\ \eta\Big(\big\{\varphi\big\}, \frac{d^2}{\ell_B^2}\Big)
\ ,\label{Ieta}
\end{align}
where $I_{c0}$ is the zero-field critical current, $\varphi=\Phi/\Phi_0=BWd/\Phi_0$ denotes the dimensionless magnetic flux (in units of the flux quantum~$\Phi_0=\pi\hbar c/e$), $d$ is the length of the wide $d\times W$ junction ($W\gg d$), and $\ell_B=\sqrt{\Phi_0/B}$ is the magnetic length. Curly brackets~$\{\varphi\}$ denote the fractional part of $\varphi$.

In the limit of short enough junctions or small enough magnetic fields, such that $d^2/\ell_B^2\ll 1$, the function~$\eta(\{\varphi\},d^2/\ell_B^2)$ in Eq.~(\ref{Ieta}) 
reduces to the known~\cite{svidzinskii} form:
\begin{equation}
  \eta_0(\{\varphi\})=\{\varphi\}(1-\{\varphi\})
\ .\label{svid1}
\end{equation}
This function for the SNS junction differs from the corresponding function for the Fraunhofer pattern in conventional Josephson tunnel (SIS) junctions,\cite{tinkham} 
for which $\eta_0(\{\varphi\})=|\sin(\pi\varphi)|/\pi$, yet these two functions share an important property: both turn zero at integer $\varphi$. 
The physical origin~\cite{svidzinskii,tinkham} of the zeros in $\eta_0(\{\varphi\})$ is also common for both the SIS and thin SNS junctions: the Josephson current density 
is a periodic function of the coordinate~$y$ along the interface with period~$\propto 1/B$ and zero mean over each full period.

In a junction of finite length~$d$, due to the electron trajectories bouncing off the side edges, 
contributions to the supercurrent from the regions near the side edges of the SNS junstion, at $y=\pm W/2$, cf. Fig.~\ref{fig_sns}, 
form differently as compared to those in the bulk region. That difference 
leads to a similar effect as an inhomogeneity of the current density, resulting in lifted zeros of the critical current~$I_c(B)$. As long as the magnetic field is small ($B\ll\Phi_0/d^2$, or equivalently $d^2/\ell_B^2\ll 1$),
such qualitative changes in $I_c(B)$ are brought by perturbative corrections to Eq.~(\ref{svid1}).

However, the effect of the edges becomes increasingly important with the increase of the magnetic field. 
For $B\gg\Phi_0/d^2$, the Josephson current is substantially determined by the nature of electron trajectories close to and bouncing off the side edges. 
For specular reflection off straight edges, regular oscillations persist. In this situation, the maxima and minima of the
oscillations of $I_c(\varphi)$ are all of the same order,
\begin{align}
I_c(B) \sim 
\frac{I_{c0}}{\varphi}\frac{\ell_B^2}{d^2}
\ .
\label{Ilargefield}
\end{align}
The critical current thus decays as $1/B^2$ with the additional factor stemming from
the second argument of the function~$\eta(\{\varphi\},d^2/\ell_B^2)$ in Eq.~(\ref{Ieta}).

Realistic side edges are rough so that the critical current acquires a random component. If this
roughness varies on a scale larger than the geometric mean of~$d$ and the electron Fermi wavelength~$\lambda_F$ in the normal layer, sample-to-sample fluctuations
of the critical current are of classical nature. Their typical amplitude,
\begin{align}
\delta I_c \sim\frac{I_{c0}}{\varphi}\frac{b_0}{d}
\ ,\label{deltaIcintro}
\end{align}
decays as $1/B$, which is slower than the decrease of the average critical current, cf. Eq.~(\ref{Ilargefield}). 
Once the fluctuation amplitude~$\delta I_c$, Eq.(\ref{deltaIc_cl}), exceeds the average critical current, it defines not only the amplitude of mesoscopic fluctuations 
but also the typical value of the critical current. The field $B_\sigma\sim\Phi_0/(b_0d)$ at which the crossover into the regime of strong fluctuations occurs depends on the amplitude of the edge roughness~$b_0$. Note that the crossover into the regime of mesoscopic fluctuations typically happens within the regime of edge-dominated transport, $B_\sigma\gg \Phi_0/d^2$.

A semiclassical theory leading to the main conclusions of this paper is organized in this manuscript as following. In the next section, 
we describe the main approximations used in this study and offer a qualitative and quantitative discussion of its results. 
In Sec.~\ref{sec:noboundaries}, we reproduce the result
of Ref.~\onlinecite{svidzinskii}, which is asymptotically accurate in the limit $B\ll \Phi_0/d^2$, by elementary means. This allows us to develop a formalism 
to treat magnetic fields of order~$\Phi_0/d^2$ and beyond, where the effects of side edge scattering become important. 
In Sec.~\ref{sec:reflection}, we derive our main results for the case of specular reflection off the edges, and 
in Sec.~\ref{sec:random}, we study the effects of the edge roughness on the $I_c(B)$ dependence. In the latter section, we also investigate the
mesoscopic fluctuations originating from edge disorder. Finally, we sum up and discuss our analysis in Sec.~\ref{sec:discussion}.

\section{Qualitative considerations and main results}

In an SNS junction (Fig.~\ref{fig_sns}) of fixed length~$d$, the crossover from short-length limit~(\ref{svid1}) to the regime of edge-dominated transport, Eq.~(\ref{Ilargefield}),
is driven by an increase of the magnetic field. We assume that the Fermi wave length~$\lambda_F$ is short enough so that at the crossover field $B\sim \Phi_0/d^2$ the cyclotron radius is still large, 
\begin{equation}
r_B\gg d
\ . \label{limit1}
\end{equation}
This condition allows for the crossover to occur before effects of quantum Hall physics\cite{hoppezuelickeschoen} set in. In addition, to simplify the theoretical consideration of the SN interfaces, 
we make the conventional~\cite{svidzinskii} assumptions about the coherence length $\xi$ and the magnetic field penetration depth~$\lambda_L$ in the superconducting leads,
\begin{align}
\lambda_F \ll (\xi,\lambda_L) \ll d.
\label{limits}
\end{align}
Conditions~(\ref{limits}) and~(\ref{limit1}) allow us to use the semi-classical approximation for the electron dynamics and to dispense with the bending of semi-classical electron trajectories, respectively. In addition, we constrain our considerations to low temperatures, $k_BT\ll \hbar v_F/d$, where the electron trajectory effects in $I_c$ are strongest ($v_F$ denotes the Fermi velocity in the normal layer). We also concentrate of ballistic junctions; in the opposite limit of diffusive SNS junctions\cite{bergeret,ivanov,bouchiat} the role of scattering off the side edges is less important. As mentioned above, we will furthermore explicitly make use of
the large aspect ratio of the wide SNS junction,
\begin{align}
W/d\gg 1\ .
\end{align}
Finally, we point out that the ``strip geometry'' of the SNS junction under consideration (Fig.~\ref{fig_sns}) is to be distinguished from a geometry of superconducting point contacts to an open normal layer, as studied in Refs.~\onlinecite{barzykinzagoskin,blanter08}.

Electron trajectories bouncing off the side edges are typically situated up to a distance~$\sim d$ from each edge. On the other hand, once flux $\Phi=BWd$ exceeds $\Phi_0$, the Josephson current density oscillates in $y$-direction with the period $\ell_B^2/d$, independent of the junction width $W$. As long as the period exceeds $d$, the bouncing trajectories weakly affect the current density distribution; the condition $\ell_B^2/d\gg d$ defines the corresponding field region $B\ll\Phi_0/d^2$, which is thus also independent of $W$.  In that region, qualitative (but quantitatively still small) effects,  such as the lifting of zeros of~$I_c(\Phi)$, become manifest only if the magnetic flux~$\Phi$ approaches an integer multiple of~$\Phi_0$.

\begin{figure}[t]
\centerline{\includegraphics[width=0.4\linewidth]{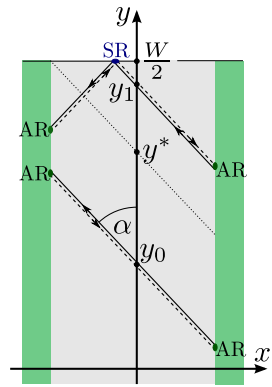}}
\caption{
(Color online) Semi-classical trajectories with fixed intersecting angle~$\alpha$ [related to momentum~$p_y$ by Eq.~(\ref{alpha_p})]
for various intersecting coordinates~$y$. For coordinates~$y_0<y^*$, cf. Eq.~(\ref{yc}), the trajectory is straight between
the superconductors at which particles (solid lines) are Andreev-reflected (AR) into holes (dashed lines). For $y_1>y^*$, trajectories involve additionally
specular reflection (SR) from the side edge.
}
\label{fig_geometry}
\end{figure}

In the limit of strong fields, $B\gg\Phi_0/d^2$, however, the scale~$\ell_B^2/d$ on which the Josephson current density oscillates along the interface line, 
is much shorter than~$d$ so that edge effects become significant. In the periods close to the edges at~$y=\pm W/2$, electron and hole trajectories that do not collide with
the side edges must then have incidence almost normal to the SN interfaces. The corresponding span of  
angles~$\alpha$, cf. Fig.~\ref{fig_geometry}, scales as~$1/B$. This leads, for collision-free trajectories, to an enhanced~$1/B^2$ decay, cf.  Eq.~(\ref{Ilargefield}). For trajectories involving reflection from edges, our calculation predicts the same enhanced decay. We find the scaling~$I_c\propto 1/B^2$ for both minima and maxima of~$I_c(\Phi)$, which thus are of same order for $B\gg\Phi_0/d^2$. 

The crossover between the limits of weak and strong fields is embodied by the dependence of the function 
$\eta(\{\varphi\},t)$ in Eq.~(\ref{Ieta}) on its second argument. It is this additional dependence that embodies the presence of a characteristic field $B\sim\Phi_0/d^2$ and the different regimes associated with the effect of the edges. In Sec.~\ref{sec:reflection}, we obtain explicit results for the asymptotic behavior of 
the function~$\eta(\{\varphi\},t)$ in the limits of $t\ll 1$ and $t\gg 1$, which correspond to magnetic fields below and above the crossover region 
($B\sim\Phi_0/d^2$). For the maxima of~$\eta(\{\varphi\},t)$ with respect to $\{\varphi\}$, we find
\begin{align}
\max\limits_{\{\varphi\}}\eta(\{\varphi\},t) &\simeq 
 \left\{
  \begin{array}{cl}
    \frac{1}{4} &\ ,\quad t\ll 1 \\
    \frac{8}{9\pi\sqrt{3}}\ t^{-1} &\ ,\quad t\gg 1
  \end{array}
 \right. 
\ .\label{result_max}
\end{align}
(Strictly speaking, the~$t\ll 1$ behavior is realized for fluxes~$\Phi$ in the range~$\Phi_0\ll \Phi\ll (W/d)\Phi_0$.) The minima of $\eta(\{\varphi\},t)$ 
with respect to $\{\varphi\}$ are finite at any $t$,
\begin{align}
\min\limits_{\{\varphi\}}\eta(\{\varphi\},t) &\simeq 
 \left\{
  \begin{array}{cl}
    \frac{f_0}{2}\ t &\ ,\quad t\ll 1 \\
    \frac{1}{9\pi\sqrt{3}}\ t^{-1} &\ ,\quad t\gg 1
  \end{array}
 \right.
\ .\label{result_min} 
\end{align}
As was already mentioned, substantial deviations from the conventional Fraunhofer pattern occur at $t\gg 1$. 
Equations ~(\ref{result_max}) and~(\ref{result_min}) summarize our results, which are fully presented in Eqs.~(\ref{maxI_low}), (\ref{minI_low}), (\ref{IExtreme}), 
and~(\ref{IExtremeLow}). The constant~$f_0$ is approximately equal to~$0.22$, cf. Eq.~(\ref{f0}). 

\begin{figure*}[t]
\centerline{\includegraphics[width=0.95\linewidth]{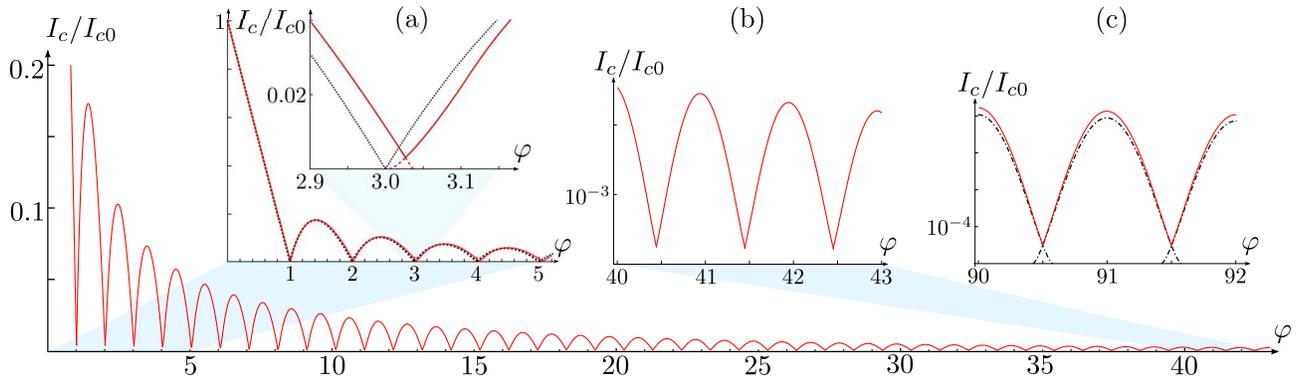}}
\caption{(Color online)
Josephson critical current~$I_c$ as function of the dimensionles magnetic flux~$\varphi=\Phi/\Phi_0$ 
under inclusion of the effects of reflection from side edges as
obtained from numerical evaluation of Eqs.~(\ref{I0a}) and~(\ref{I1a}) assuming an aspect ratio
of $W/d=20$. (a) At low flux, $\varphi\ll W/d$, scattering off side edges
leads to merely small alterations of the result~(\ref{svidzinskiJ_current}), represented
by the dotted line, of Ref.~\onlinecite{svidzinskii}. Maxima and minima are given by Eqs.~(\ref{maxI_low}) and~(\ref{minI_low}). 
\emph{Inset}: Zoom into an interval close to a minimum, showing the non-zero value of~$\min I_c$ and the slight shift to the right due to scattering effects.
Red dashed lines indicate the continuation of henceforth local maxima of~$I(\chi_0,\varphi)$ as a function of~$\chi_0$.
(b) For larger flux~$\varphi> W/d$, because of side-edge effects, the (non-zero) minima of~$I_c(\varphi)$ 
have shifted to half-integer values of $\varphi$ and scale, like the maxima, as~$1/\varphi^2$, cf. Eqs.~(\ref{IExtreme})
and~(\ref{IExtremeLow}). (c) Asymptotic ``bell-shaped'' curve (dash-dotted line) of Eq.~(\ref{IMax}), which is valid in the limit~$\varphi\gg W/d\gg 1$.
}
\label{fig_fraunhofer}
\end{figure*}

The hallmarks of the ``modified Fraunhofer'' pattern~(\ref{Ieta}) are displayed graphically in Fig.~\ref{fig_fraunhofer}. 
Despite qualitative and quantitative changes to $I_c(\Phi)$ as compared to the conventional Fraunhofer pattern, the derivative $\mathrm{d}I_c/\mathrm{d}\Phi$ remains discontinuous at the current minima. 

A more subtle observation from the theory presented in the next sections is the ``creep'' 
of the minima of~$I_c(\Phi)$, shifting them away from integer multiples~$\Phi=n\Phi_0$. Linear in~$B$ for $B\ll\Phi_0/d^2$, this shift
saturates at the characteristic field, $B\sim\Phi_0/d^2$, to half a period. For large fields, $B\gg\Phi_0/d^2$, minima are thus situated at~$\Phi=(n+1/2)\Phi_0$ while the maxima
have shifted to integer multiples of~$\Phi_0$. We may interpret this shift as a reflection of a crossover from the interference pattern of a (wide) single slit
at low fields to an effective double-slit interference pattern at larger fields, at which the Josephson transport is dominated by the two side edges.
We note that in the situation of edge transport in quantum spin Hall interfaces, for a similar reason, the periodicity in the magnetic interference pattern 
resembles that of a double slit as well.\cite{yacoby14,beenakker15}

Randomness in scattering off the edges leads to mesoscopic, sample-to-sample fluctuations $\delta I_c$ of the critical current. As long as $\delta I_c$
remains small compared to the typical average critical current, Eq.~(\ref{Ieta}), the latter effectively provides the description of the experimental
observable. For a small-amplitude ``classical'' edge randomness, $\lambda_F\ll b_0\ll d$, mesoscopic fluctuations exceed the average current at a magnetic field~$B_\sigma\sim\Phi_0/(b_0d)\gg\Phi_0/d^2$, see Eq.~(\ref{phic_cl}). In higher fields, $B\gtrsim B_\sigma$, $\delta I_c$, cf. Eqs.~(\ref{deltaIcintro}) and~(\ref{deltaIc_cl}), defines both amplitude of mesoscopic fluctuations and the typical value of the critical current. The corresponding estimates for the diffractive edge scattering, corresponding
to edge disorder with correlation length $\sim\lambda_F$, are given by Eqs.~(\ref{deltaIc}) and (\ref{phic}) in Sec.~\ref{sec:random}.

\section{SNS junction without side edges}
\label{sec:noboundaries}

In this section, we rederive the formula\cite{svidzinskii} for the Josephson current through a long and wide
SNS junction as depicted in Fig.~\ref{fig_sns} neglecting any effects due to the edges
in $y$-direction. The derivation
we present here is elementary and, unlike the one provided in Ref.~\onlinecite{svidzinskii}, does not
involve methods based on Matsubara Green's functions. Realistic boundary conditions, which involve specular
reflection, will be studied in Sec.~\ref{sec:reflection} within the semi-classical framework presented in this section.

\subsection{Josephson transport at zero magnetic field}

We obtain the energy spectrum of the SNS junction in Fig.~\ref{fig_sns} under consideration
by solving the Boguliobov--de-Gennes equations:\cite{deGennes}
\begin{align}
\left[
-\frac{1}{2m}\nabla^2-\varepsilon_F
\right] \psi_e + \Delta(x)\psi_h &= E\psi_e
\nonumber\\
\left[
\frac{1}{2m}\nabla^2+\varepsilon_F
\right] \psi_h + \Delta^*(x)\psi_e &= E\psi_h
\ . \label{bdg}
\end{align}
(If not stated otherwise, we set~$\hbar=1$ throughout the text.) In these equations, $m$ denotes the electronic (effective) mass, $\varepsilon_F$ is the Fermi energy,
$\nabla=(\partial_x,\partial_y)$, and~$E$ the eigenenergy of the state represented by the particle
and hole-type wave functions~$\psi_e$ and $\psi_h$. The superconducting order parameter, in a symmetric gauge, has the form
\begin{align}
\Delta(x) &= |\Delta| \Theta(|x|-d/2)\mathrm{e}^{\mathrm{i}\,\sgn(x)\chi/2}
\ ,
\end{align}
where $\Theta$ denotes the Heaviside step function and
\begin{align}
\chi &= \chi_R-\chi_L
\end{align}
the phase difference between the right and left superconductors. In the conventional BCS limit, $|\Delta|\ll\varepsilon_F$ or $\lambda_F\ll\xi$,
the semi-classical approximation yields the energy levels below the superconducting gap~$|\Delta|$ and the corresponding 
wave functions $\psi_e$ and $\psi_h$ for the important domains of momentum. In the normal region, these read
\begin{align}
\psi_{e}^{n,\pm} &= 
\frac{1}{\sqrt{Wd}} 
\mathrm{e}^{-\frac{\mathrm{i}\chi}{4}}
\mathrm{e}^{
\pm\mathrm{i}(\tilde{p}_F + E_{n,\pm}/\tilde{v}_F)x
}
\sin\left[p_y \left(y+\frac{W}{2}\right)\right] \ ,\nonumber\\
\psi_{h}^{n,\pm} &= \frac{\mp(-1)^{n}}{\sqrt{Wd}}
\mathrm{e}^{\frac{\mathrm{i}\chi}{4}}
\mathrm{e}^{
\pm\mathrm{i}(\tilde{p}_F - E_{n,\pm}/\tilde{v}_F)x
} \sin\left[p_y \left(y+\frac{W}{2}\right)\right]
\ .\label{semiclassicalwavefunctions}
\end{align}
The quantities $\tilde{p}_F=p_F\sin\alpha$ and $\tilde{v}_F=v_F\sin\alpha$ denote a ``reduced'' Fermi momentum and
a ``reduced'' Fermi velocity, respectively, with the angle~$\alpha$
defined by
\begin{align}
\cos\alpha &= \frac{p_y}{p_F}
\ .\label{alpha_p}
\end{align}
The eigenenergies~$E_{n,\pm}$, where index $n$ is an integer and~$\pm$ distinguishes 
the sectors with $p_x>0$ and~$p_x<0$, are given by
\begin{align}
\pm \frac{2 E_{n,\pm}\,\tilde{d}}{v_F} &= 2\pi\hbar\left(n+\tfrac{1}{2}\right)+\hbar\chi
\ ,\label{quantization}
\end{align}
restoring Planck's constant~$\hbar$ for a moment.
In this formula,
\begin{align}
\tilde{d} = \frac{d}{\sin\alpha}
\label{deff}
\end{align}
is the length of a semi-classical trajectory cutting
the~$y$-axis at the angle~$\alpha$, cf. Fig.~\ref{fig_geometry}.
 
We should interpret Eq.~(\ref{quantization}) 
as semi-classical Bohr-Sommerfeld quantization rule for a particle-hole pair counter-propagating
along a trajectory of length~$\tilde{d}$, cf. also the related discussion in Ref.~\onlinecite{abrikosov}. This interpretation will also allow 
for the generalization to $y$-dependent phase differences~$\chi$ and semi-classical trajectories
involving scattering off side edges. For a specific trajectory,
the phase difference~$\chi$ that enters the quantization rule~(\ref{quantization}) 
will be determined by the \emph{local} phases of the superconducting condensate at the points of Andreev reflection.

The semi-classical approximation introduced above is valid as long as
the wave number~$k_x$ in~$x$-direction is much larger than~$1/d$.
From Eq.~(\ref{semiclassicalwavefunctions}), we infer the
effective wave numbers of the Andreev states,
\begin{align}
k_x &= p_F\sin\alpha\pm\frac{E_{n,\pm}}{v_F\sin\alpha}
\ .\label{kx}
\end{align}
The bounds of the Andreev spectrum are given by~$\pm|\Delta|$.
We thus find that as long as $\sin^2\alpha\gg |\Delta|/\varepsilon_F$,
the effective wave number is of order~$p_F$, \emph{i.e.}, $\gg 1/d$, and
the  semi-classical approximation is thus valid. Its validity breaks
down for momenta~$p_y$ very close to~$\pm p_F$ such that $1-(p_y/p_F)^2\lesssim|\Delta|/\varepsilon_F$. 
These momenta, however, occupy only a small phase space domain that
is unimportant for the Josephson current.

\subsubsection{Josephson current}

With the knowledge of the wave functions~(\ref{semiclassicalwavefunctions}), 
we obtain the Josephson current as a function of the phase difference~$\chi$ using
\begin{widetext}
\begin{align}
J(\chi)
&= \frac{-e}{4m}\int\limits_{-W/2}^{W/2}\mathrm{d}y\int\limits_{-p_F}^{p_F}\frac{\mathrm{d}p_y}{2\pi/W}
\sum_{n,\varrho=\pm}\ 
\mathrm{Im}
\Big\{
f(E_{n\varrho}) \big[\psi_e^{n\varrho}\big]^*\partial_x \psi_e^{n\varrho}
+ [1-f(E_{n\varrho})] \psi_h^{n\varrho}\partial_x \big[\psi_h^{n\varrho}\big]^*
\Big\}\Big|_{x=0}
\ , \label{bdg_current}
\end{align}
\end{widetext}
where $f(\varepsilon)=[\exp(\varepsilon/k_BT)+1]^{-1}$ is the Fermi distribution function.
Keeping the temperature~$T$ non-zero provides a natural regularization of the sum over eigenlevels~$n$, and allows one to avoid the appearance of spurious terms~\cite{kulik} associated with the non-analyticity of the density of states at energy $E=|\Delta|$. Following the method detailed in Refs.~[\onlinecite{svidzinskii}] and [\onlinecite{ishii}], we convert the right-hand side of Eq.~(\ref{bdg_current}) into a sum over Matsubara frequencies.  It becomes evident then, that the leading contribution to the current comes from the energy interval controlled by the largest of the two scales, $T$ and $\hbar v_F/d$ (both are small compared to $|\Delta|$).
Taking the zero-temperature limit, we find a sawtooth-shaped current function,
\begin{align}
J(\chi) &= I_{c0}
\Bigg\{
\frac{\chi}{\pi} - 
\sum_{j=0}^{\infty}
\Big[
\Theta\left(
\frac{\chi}{\pi}-(2j+1)
\right) \nonumber\\
&\qquad\qquad\qquad\qquad -\Theta\left(
-\frac{\chi}{\pi}-(2j+1)
\right)
\Big]
\Bigg\}
\ ,\label{IZeroB}
\end{align}
where
\begin{align}
I_{c0} &= 
     \frac{ev_F}{d} \frac{p_FW}{4}
\label{Ic}
\end{align}
is the Josephson critical current at zero external field. 
The first factor in Eq.~(\ref{Ic}) corresponds to the current
in a one-dimensional link while the second factor reflects
the various transverse channels in the two-dimensional junction.
\bigskip

\subsubsection{Long-wavelength phase variations}

In order to generalize formula~(\ref{IZeroB}) to non-zero magnetic fields, let us
address the situation in which the phases~$\chi_R$ and~$\chi_L$ of the superconducting
condensates at the interfaces vary ``slowly'' as a function of~$y$ such that it is possible
to adiabatically decouple the motion in $x$ and $y$-directions.

The wave functions~$\psi_{e/h}^{n,\pm}$,
Eq.~(\ref{semiclassicalwavefunctions}), which enter the general current formula~(\ref{bdg_current}), 
oscillate at the length scale~$p_y^{-1}$. As this length is typically of the order of the Fermi wavelength, $p_y^{-1}\sim\lambda_F$,
the notion of ``slow'' variations implies variations on a much larger scale~$\ell\gg\lambda_F$ over which 
the $y$-dependence of the wave functions is effectively a constant as the fast oscillations on the
scale~$\lambda_F$ vanish on average. In this situation,
the formula~(\ref{bdg_current}) for Josephson current is effectively reduced to
the simpler form
\begin{align}
I(\chi_0)
&= \frac{2}{p_FW}\int\limits_{-W/2}^{W/2}\mathrm{d}y\int\limits_{-p_F}^{p_F} \frac{\mathrm{d}p_y}{2\pi}
   \ \frac{d}{\tilde{d}}\ J\big[\chi_0 + \delta\chi(y,p_y)\big]
   \ ,
\label{IGeneral}
\end{align}
where $J(\chi)$ is given by Eq.~(\ref{IZeroB}), $\chi_0$ denotes the average phase difference along the SNS interface, and~$\delta\chi(y,p_y)$ 
is a variation that is slow in the sense discussed above. The effective length $\tilde{d}$ of the semi-classical trajectories
depends on momentum~$p_y$ according to Eqs.~(\ref{deff}) and~(\ref{alpha_p}).

Intuitively, let us think of the integral in Eq.~(\ref{IGeneral}), which is an integral over classical phase space~$(y,p_y)$, as
the sum over current contributions from all semi-classical trajectories that connect the two superconductors
through the normal region. In fact, each phase space point~$(y,p_y)$ determines
the unique trajectory crossing the $y$-axis at point~$y$ with the angle~$\alpha$ given by Eq.~(\ref{alpha_p}), cf. also Fig.~\ref{fig_geometry}.

\subsection{Non-zero magnetic field}

Let us now address the situation of a non-zero perpendicular magnetic field~$B$,
assumed to be constant in the normal layer and to fall sharply to zero at the boundaries to the superconducting regions.
This assumption corresponds to a short London penetration depth in the sense of $\lambda_L\ll\min(d, \ell_B^2/d)$.
Following Ref.~\onlinecite{svidzinskii}, we choose a gauge for
the vector potential such that $\mathbf{A}=A_y\mathbf{e}_y$ with
\begin{align}
A_y &=
\left\{
\begin{array}{cl}
-Bx &\ ,\ -\dfrac{d}{2} < x < \dfrac{d}{2} \ ,\\
-\frac{1}{2}Bd\ \sgn x & \ ,\ |x|>\dfrac{d}{2} \ .
\end{array}
\right.
\end{align}
The condition of zero screening current in the bulk superconductor and the limit of~$\lambda_L\rightarrow 0$
require the superconducting phase at the interfaces ($x=\pm d/2$) to become functions of~$y$,
\begin{align}
\chi_{R/L} &=
\pm\frac{1}{2}\Big(
\chi_0 - \frac{2\pi \varphi y}{W}
\Big)
\ ,
\label{chi_y}
\end{align}
where $\varphi$ measures the magnetic flux~$\Phi=BdW$ penetrating the normal layer
in units of the flux quantum,
\begin{align}
\varphi &= \frac{\Phi}{\Phi_0}
\ .\label{defn}
\end{align}

In the situation of not too large magnetic fields, such that the cyclotron radius~$r_B$
remains much larger than~$d$, the semi-classical trajectories may be assumed to be straight
lines rather than circular orbits. In this approximation, used in Ref.~\onlinecite{svidzinskii},
the magnetic field enters the formalism only through the phase dependence on~$y$, Eq.~(\ref{chi_y}),
or, speaking in gauge-invariant terms, through the total flux in the normal layer.
In the present geometry, cf. Fig.~\ref{fig_geometry}, and in the notation of Eq.~(\ref{IGeneral}),
we find that
\begin{align}
\delta\chi &= - \frac{2\pi\varphi y}{W}
\ ,
\label{phasedifference_0}
\end{align}
which, in the limit $\ell_B\gg\lambda_F$, constitutes an indeed slow spatial variation of the phase difference.

Inserting Eq.~(\ref{phasedifference_0}) into Eq.~(\ref{IGeneral})
and carrying out the integration for $-\pi<\chi_0<\pi$, we obtain
\begin{align}
I(\chi_0,\varphi)
&= \frac{I_{c0}}{\varphi}
   \left\{
   \begin{array}{cl}
     \kappa\big(1-\{\varphi\},\chi_0/\pi\big) &\quad \textnormal{for $\lfloor\varphi\rfloor$ even} \\
     -\kappa\big(\{\varphi\},\chi_0/\pi\big) &\quad \textnormal{for $\lfloor\varphi\rfloor$ odd}
   \end{array}
   \right.
\label{svidzinskii_intermediate}   
\end{align}
where $\lfloor\varphi\rfloor$ denotes the integer
part of~$\varphi$, such that $\varphi=\lfloor\varphi\rfloor+\{\varphi\}$,
and we have defined
\begin{align}
\kappa(u,v) &= \frac{1}{2}
\left(
|u+v|-|u-v|-2uv
\right)
\ . \label{kappa}
\end{align}
For given~$\varphi$, the Josephson critical current is
defined to be the maximum of~$I(\chi_0,\varphi)$ with respect
to~$\chi_0$,
\begin{align}
I_c(\varphi) &=
\max\limits_{\chi_0} I(\chi_0,\varphi)
\ , \label{IcDef}
\end{align}
and is easily found using Eqs.~(\ref{svidzinskii_intermediate}) and~(\ref{kappa}).
This leads to a rather simple formula for~$I_c$ as a function
of~$\varphi=\Phi/\Phi_0$,
\begin{align}
I_c(\varphi)
&= \frac{I_{c0}}{\varphi}\ \{\varphi\}(1-\{\varphi\})
\ ,\label{svidzinskiJ_current}
\end{align}
cf. Eq.~(\ref{svid1}).
Formula~(\ref{svidzinskiJ_current}) was first reported in Ref.~\onlinecite{svidzinskii}. It is plotted
in Fig.~\ref{fig_fraunhofer}(a) in comparison with the results including
scattering off side edges, which we derive in the next section.

\section{Effects of scattering off side edges}
\label{sec:reflection}

In this section, we generalize the earlier result~(\ref{svidzinskiJ_current}) of Ref.~\onlinecite{svidzinskii} by taking into account
reflections from the side edges. We assume here clean edges from which reflection is specular.
Random edges will be addressed in Sec.~\ref{sec:random}. In the limit of a small magnetic field, $B\ll\Phi_0/d^2$, 
equivalently $\varphi\ll W/d$, side edge
effects are perturbative corrections but become dominant in the opposite limit~$B\gg\Phi_0/d^2$.

\subsection{Semi-classical geometric picture}

We adopt the semi-classical picture of straight trajectories, assuming small enough magnetic fields
that allow us to neglect the curvature of the orbits. If the junction's width~$W$ and hence the ratio of~$W$ to its length~$d$
are finite, a subset of the semi-classical trajectories has to involve one or multiple reflections from the side edges.
In a first step, we should classify the total set of trajectories into subsets of trajectories with the same number of
reflections from the side edges. Since each trajectory corresponds to a point in the classical phase space~$(y,p_y)$,
cf. Eq~(\ref{IGeneral}), this amounts to subdividing the classical phase space into different regions characterized
by the integer number of side reflections.

Let us restrict ourselves to the positive quadrant, $y>0$ and $p_y>0$, for the moment. As will become
evident shortly, trajectories with more than one reflection contribute only subleading corrections to the current.
It is thus sufficient to restrict ourselves to momenta~$p_y$ corresponding to angles~$\alpha$ with $\tan\alpha > d/W$, cf. Fig.~\ref{fig_geometry},
\emph{i.e.},
\begin{align}
p_y < \frac{p_F}{\sqrt{1+(d/W)^2}}
\ .
\end{align}
This inequality, in combination with Eq.~(\ref{IGeneral}), already shows that neglecting trajectories with more reflections merely adds up to a small 
error of order $(d/W)^2$ in the calculation of the current. Intuitively, we understand this quadratic smallness as trajectories
with more than one reflection appear only in a small angular domain, $\alpha\lesssim d/W$, and 
furthermore contribute a much smaller current because of their large length~$\sim W$.

For fixed~$p_y$, and hence fixed angle $\alpha$, we find that a trajectory features no reflections
as long as $y<y^*(\alpha)$ with
\begin{align}
y^*(\alpha) &= \frac{W}{2}
	\left(
	 1- \frac{d}{W}\frac{1}{\tan\alpha}
	\right)
\ .
\label{yc}
\end{align}
Trajectories with $y^*(\alpha)<y<W/2$ feature exactly one reflection, cf. Fig.~\ref{fig_geometry}. 

The phase difference~$\chi_0+\delta\chi(y,p_y)$, which enters the semi-classical
quantization rule~(\ref{quantization}), depends crucially on whether the trajectory
is straight or involves reflection from a side edge. In the former case, \emph{i.e.}, for $y<y^*(\alpha)$,
Eq.~(\ref{phasedifference_0}) still applies, whereas for trajectories featuring a single reflection
the phase difference~$\chi_0+\delta\chi$ becomes independent of~$y$:
\begin{align}
\delta\chi &= - \frac{2\pi\varphi}{W}\ y^*(\alpha)
\label{phasedifference_1}
\end{align}
for $y^*(\alpha)<y<W/2$. Through~$\alpha$, the phase difference~$\delta\chi$ depends on $p_y$ though, cf. Eq.~(\ref{alpha_p}).
The different dependence of the phase on the phase space coordinates is what distinguishes
semi-classical trajectories with and without reflections.

Equation~(\ref{IGeneral}) now allows us to immediately write down an expression for the current contribution
due to semi-classical states $(y,p_y)$ with $y>0$ and $p_y>0$. The other phase space quadrants
lead to analogous contributions as we note that spatial reflection $y\leftrightarrow -y$ converts $\delta\chi$ to $-\delta\chi$ while
momentum reflection $p_y\leftrightarrow -p_y$ leaves $\delta\chi$, and hence the current, invariant.

\subsection{Current formula}

As a result of the discussion of the preceding section, we represent
the total zero-temperature Josephson current through the SNS junction, Eq.~(\ref{IGeneral}), as the sum of 
contributions from semi-classical trajectories with zero reflections
and those with one reflection,
\begin{align}
I(\chi_0,\varphi) &= I_0(\chi_0,\varphi)+I_1(\chi_0,\varphi)
\ . \label{Isum}
\end{align}
Technically, they appear as we divide the $y$-integrals in Eq.~(\ref{IGeneral})
at~$y=y^*$, Eq.~(\ref{yc}), with the domain $y<y^*$ yielding~$I_0$ and $y>y^*$ yielding~$I_1$.

Evaluating the current contributions~$I_0$ and~$I_1$ separately, we find
\begin{widetext}
\begin{align}
I_0(\chi_0,\varphi)
&= \frac{2I_{c0}}{\pi}
 \left\{\frac{\chi_0}{2}-\frac{d}{W}\frac{\chi_0}{\pi}
 - \frac{d}{W} \sum_{j=0}^\infty
 \left[
        \bar{F}_0\left(
  	    \frac{W/d}{\varphi}\Big(\frac{\chi_0}{\pi}-(2j+1)+\varphi\Big)
        \right)
          -
        \bar{F}_0\left(
  	    \frac{W/d}{\varphi}\Big(-\frac{\chi_0}{\pi}-(2j+1)+\varphi\Big)
        \right)        
  \right]
 \right\} 
\ , \label{I0a} \\
I_1(\chi_0,\varphi)
&= \frac{2I_{c0}}{\pi}
 \left\{\frac{d}{W}\frac{\chi_0}{\pi}
 - \frac{d}{W} \sum_{j=0}^\infty
 \left[
        \bar{F}_1\left(
  	    \frac{W/d}{\varphi}\Big(\frac{\chi_0}{\pi}-(2j+1)+\varphi\Big)
        \right)
          -
        \bar{F}_1\left(
  	    \frac{W/d}{\varphi}\Big(-\frac{\chi_0}{\pi}-(2j+1)+\varphi\Big)
        \right)        
  \right]
 \right\} 
 \label{I1a}
\end{align}
\end{widetext}
with $\bar{F}_l(u)=\Theta(u)F_l(u)$ for $l=1,2$ and
\begin{align}
F_0(u) &= u\arctan u \ , 
\label{F0}\\
F_1(u) &= u^2/(1+u^2)\ .
\label{F1}
\end{align}
The Josephson current~(\ref{Isum}) together with Eqs.~(\ref{I0a}) and~(\ref{I1a})
is an odd function of the phase, $I(\chi_0,\varphi)=-I(-\chi_0,\varphi)$,
which also follows immediately from the assumed reflection symmetries with respect to $x$ and $y$ axes (and 
spin-rotational symmetry).\cite{flensberg15}

Calculating the Josephson current numerically for a given ratio~$W/d$ by means of 
the above formulas, we find for the Josephson critical current~$I_c$, Eq.~(\ref{IcDef}), as a function of
flux~$\varphi$ the ``modified Fraunhofer pattern'' shown in Fig.~\ref{fig_fraunhofer}.
In the following, we are going to study the function~$I_c(\varphi)$ analytically in the two
limiting cases~$\varphi\ll W/d$ and~$\varphi\gg W/d$.

\subsection{Josephson current at small field $B\ll \Phi_0/d^2$}

In the small field regime, $B\ll \Phi_0/d^2$ or $\varphi\ll W/d$, the typical value~$I_{c0}/\varphi$
of the Josephson critical current in the absence of edge effects, cf. Eq.~(\ref{svidzinskiJ_current}), 
is much larger than corrections due to scattering off side edges, whose contribution to 
the current is smaller by a factor of~$W/d\gg 1$. However, for~$\varphi$ close
to the minima of~$I_c(\varphi)$, which were zero in Eq.~(\ref{svidzinskiJ_current}), 
effects of reflection from side edges constitute the leading order contribution to
the current.

\subsubsection{Maxima of the critical current}

The maxima of the Josephson critical current~$I_c(\varphi)$ in
the regime $\varphi\ll 1$ occur at values~$\varphi$ for which
the analysis of Ref.~\onlinecite{svidzinskii}, reproduced in Sec.~\ref{sec:noboundaries}, applies. 
For $\varphi\gtrsim 1$, the maxima of $I_c(\varphi)$ are, according to Eq.~(\ref{svidzinskiJ_current}),
given by
\begin{align}
\max I_c(\varphi) \simeq \frac{I_{c0}}{4\varphi}
\ .\label{maxI_low}
\end{align}
This formula corresponds to the upper line in Eq.~(\ref{result_max}).

\begin{figure}[t]
\centerline{\includegraphics[width=\linewidth]{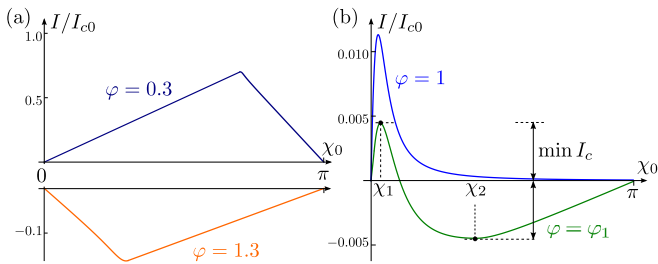}}
\caption{(Color online)
Josephson current as a function of average phase difference phase difference~$\chi_0$ for
various magnetic fluxes~$\varphi$ in the regime~$\varphi\ll W/d$ for here~$W/d=20$. 
(a) The current at fluxes between integer multiples of the flux quantum 
is adequately described by Eq.~(\ref{svidzinskii_intermediate}) as corrections
in $d/W$ merely lead to slightly rounding the peaks. (b) \emph{Upper curve}: Non-zero
Josephson current at $\varphi=1$, cf. Eq.~(\ref{IOddInteger}). \emph{Lower curve}: The critical
current~$I_c$ becomes minimal at slightly greater~$\varphi=\varphi_1\approx 1.007$, cf. Eq.~(\ref{minima})
and the discussion in Sec.~\ref{ssec:minima_low}.}
\label{fig_lowflux}
\end{figure}

Reflection corrections to $I_c(\varphi)$ close to its maxima and also for~$\varphi\ll 1$ 
are small. For fluxes~$\varphi\ll W/d$, these corrections merely lead to
slight roundings of the peaks in the $\chi_0$-dependence of~$I(\chi_0,\varphi)$, cf. Fig.~\ref{fig_lowflux}(a).
These roundings induce only subleading corrections of relative order~$d/W$ to Eq.~(\ref{maxI_low}).

\subsubsection{Critical current at integer multiples of~$\Phi_0$}

According to Eq.~(\ref{svidzinskiJ_current}), if we ignore effects of reflection from side edges,
there is no net Josephson current whenever the magnetic flux~$\Phi$ equals an integer multiple of the flux quantum~$\Phi_0$.
Including these effects, we find with Eqs.~(\ref{I0a}) and~(\ref{I1a}),
that the Josephson current for positive integer~$\varphi=\Phi/\Phi_0$ is non-zero, although small in the
aspect ratio~$d/W$. This smallness reflects its origin at the edges.
For odd integer, $\varphi=2n+1$ with~$n\ll W/d$, we find
\begin{align}
I(\chi_0,\varphi=2n+1)
&=
\frac{I_{c0}}{W/d}
\Big[
\frac{W/d}{\varphi}\frac{\chi_0}{\pi}
-
\frac{2}{\pi}F\Big(
\frac{W/d}{\varphi}\frac{\chi_0}{\pi}
\Big)
\Big]
\ ,\label{IOddInteger}
\end{align}
which is plotted in Fig.~\ref{fig_lowflux}(b) for $\varphi=1$.
Formula~(\ref{IOddInteger}) is written for $0<\chi_0<\pi$, and we have introduced the function
\begin{align}
F(u) &= F_0(u)+F_1(u)
\ ,
\end{align}
cf. Eqs.~(\ref{F0}) and~(\ref{F1}).
For even integer, $\varphi=2n$, an analogous expression is obtained
with $\chi_0$ replaced by~$\pi-\chi_0$ and with the opposite overall sign.
For this reason, in order to study the Josephson critical current, it is sufficient 
to consider fluxes~$\varphi$ 
at and close to odd integers.

Inspecting Eq.~(\ref{IOddInteger}), we
find that the current is maximal at average phase difference 
$\chi_0= u_0\pi(d/W)\varphi$ with $u_0\approx 0.55$ defined by $F'(u_0)=\pi/2$.
The maximal current at integer~$\varphi\ll W/d$ is
\begin{align}
I_c(\varphi\in\mathbb{N}) &\simeq f_0 \frac{I_{c0}}{W/d}
\ ,\label{IcInteger}
\end{align}
where the numerical coefficient~$f_0$ is defined by
\begin{align}
f_0 &= u_0-\frac{2F(u_0)}{\pi} \approx 0.22
\ .\label{f0}
\end{align}
Remarkably, the critical current~$I_c$ at integer~$\varphi$ is non-zero, which
contrasts the prediction of Eq.~(\ref{svidzinskiJ_current}) that has been derived 
neglecting reflection effects. We note that the current~$I_c$ 
at integer~$\varphi$ is (to leading order) independent of the value of the magnetic flux~$\varphi$.

\subsubsection{Minima of the critical current}

\label{ssec:minima_low}

In the preceding section, we saw that due to effects of reflection off
side edges the zeros in the Fraunhofer-type pattern~(\ref{svidzinskiJ_current})
are lifted to small but non-zero values, cf. Eq.~(\ref{IcInteger}). 
The dependence of current $I(\chi_0,\varphi)$ on $\chi_0>0$ 
at $\varphi=1$ is presented by the upper curve in Fig.~\ref{fig_lowflux}(b). 
Fixing $\varphi$ at a slightly higher value additionally brings a bulk contribution to~$I(\chi_0,\varphi)$, as described by Eq.~(\ref{svidzinskiJ_current}). 
That contribution is negative, cf. the lower panel in Fig.~\ref{fig_lowflux}(a), and therefore results
in a highly non-monotonic current dependence on~$\chi_0$, attaining both positive and negative values. 
With the increase of flux, the maximum at~$\chi_1$, see Fig.~\ref{fig_lowflux}(b), is decreasing, while 
the absolute value of the minimum at~$\chi_2$ is increasing. For a specific value of flux $\varphi=\varphi_1$, 
we reach a point at which $I(\chi_1,\varphi_1)=|I(\chi_2,\varphi_1)|$. That flux~$\varphi_1$ 
corresponds to the minimal value of the critical current~$I_c$, Eq.~(\ref{IcDef}), cf. the lower curve in Fig.~\ref{fig_lowflux}(b). 
The value~$\varphi_1$ only slightly exceeds $1$ by an amount $\sim d/W$. 
A similar picture is true for each of the critical current minima in the weak-field regime~$B\ll \Phi_0/d^2$.

An analytical investigation of these effects based on Eqs.~(\ref{I0a}) and~(\ref{I1a}) shows that
the minima of the Josephson critical current~$I_c(\varphi)$ occur at fluxes
\begin{align}
\varphi_m &\simeq \left(1+\frac{f_0}{2W/d}\right)m
\ ,
\label{minima}
\end{align}
with $m$ being a non-zero integer, and amount to
\begin{align}
\min I_c(\varphi) &\simeq \frac{f_0}{2} \frac{I_{c0}}{W/d}
\label{minI_low}
\end{align}
with the numerical coefficient~$f_0$ given by Eq.~(\ref{f0}).

On top of being non-zero, the values of the minima, Eq.~(\ref{minI_low}), 
are notably independent of~$\varphi$. From this result,
we obtain the upper line of Eq.~(\ref{result_min}).
Equation~(\ref{minima}) furthermore 
predicts a ``period'' of the ``modified Fraunhofer'' pattern that is slightly increased from one flux quantum
by the geometry-dependent amount of $f_0 d/2W$ flux quanta.

We observe that, being independent of~$B$ to leading order, the minima of the Josephson critical current, 
Eq.~(\ref{minI_low}), do not scale with~$B$ in parallel with the maxima, which according to Eq.~(\ref{maxI_low})
fall as~$\propto 1/B$. Comparing the two mentioned equations, we immediately understand
that the results in this section break down at fluxes~$\varphi\sim W/d$. For $\varphi\gg W/d$,
we have to expect a different scaling behavior, which we will study in the next section.

\subsection{Josephson current at large field $B\gg \Phi_0/d^2$}

When the magnetic field is large, $B\gg \Phi_0/d^2$ or equivalently $\varphi\gg W/d$, the Josephson current
in the bulk averages to zero for any~$\{\varphi\}$ and finite contributions arise inside a strip of width~$d$
at the side edges only. Here, the question whether a particle propagating along
a semi-classical trajectory undergoes reflection from the side edges or not is essential.
As a result, Eq.~(\ref{svidzinskiJ_current}), even in combination with
the small alterations discussed in the preceding section, is no longer applicable. We emphasize, though, that the magnetic field
is still assumed small enough such that the cyclotron radius is large, $r_B \gg d$, 
so that the system is not yet in the quantum Hall regime of skipping Andreev orbits at the edge.\cite{hoppezuelickeschoen}

In the limit~$\varphi\gg W/d$, Eqs.~(\ref{I0a}) and~(\ref{I1a}) allow us to extract
an analytical expression for the Josephson current. 
Let us split from the dimensionless flux~$\varphi$ the even-integer part,
\begin{align}
\varphi = 2(n + \nu)
\label{NSplit}
\end{align}
where $n\gg 1$ is an integer and $0<\nu<1$. Inserting Eq.~(\ref{NSplit}) into Eqs.~(\ref{I0a}) and~(\ref{I1a})
and using the Euler-Maclaurin formula to evaluate the sums in~$j$ between~$0$ and~$n-1$,
we find that
\begin{widetext}
\begin{align}
I_0(\chi_0,\varphi) \simeq I_1(\chi_0,\varphi)
\simeq
\frac{2}{\pi}\frac{I_{c0}}{\varphi}
\frac{\ell_B^2}{d^2}
\Bigg\{&
\frac{\chi_0}{3\pi}
\Big[\Big(\frac{\chi_0}{\pi}\Big)^2 - (1-12\nu^2)\Big]
-\Theta\Big(\frac{\chi_0}{\pi}-(1-2\nu)\Big)
\Big[\frac{\chi_0}{\pi}-(1-2\nu)\Big]^2
\nonumber\\
& +
\Theta\Big(-\Big[\frac{\chi_0}{\pi}+(1-2\nu)\Big]\Big)
\Big[\frac{\chi_0}{\pi}+(1-2\nu)\Big]^2
\Bigg\}
\ .
\label{ILargeFlux}
\end{align}
\end{widetext}
Interestingly, in the large-field limit, one half of the total
Josephson current~$I=I_1+I_2$ is due to straight trajectories and one half due
to trajectories with one reflection from the side edges. The Josephson current~$I\propto\ell_B^2/\varphi$
falls as $1/B^2$, in contrast to the $1/B$-behavior in the usual Fraunhofer pattern.

\begin{figure}[t]
\centerline{\includegraphics[width=0.65\linewidth]{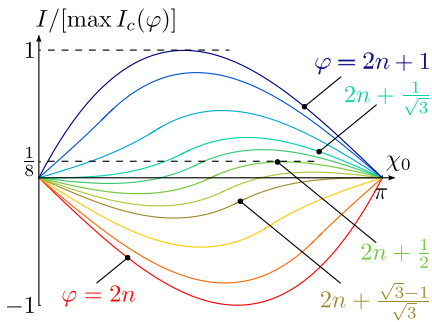}}
\caption{(Color online)
Josephson current~$I$ as a function of the average phase difference~$\chi_0$
for various magnetic fluxes between $\varphi=2n$ and~$\varphi=2n+1$,
where $n$ is an integer number $\gg W/d$. Current is measured in units of $\max I_c(\varphi)$ as given in Eq.~(\ref{IExtreme}).
}
\label{fig_lf}
\end{figure}

Figure~\ref{fig_lf} shows the Josephson current
as a function of the average phase difference~$\chi_0$ for various magnetic fluxes between two neighboring integer 
multiples of the flux quantum. It illustrates that the critical current $I_c$, Eq.~(\ref{IcDef}), is largest for integer multiples of the flux quantum
and minimal at half-integer multiples. This constitutes a ``shift'' by half a flux quantum from the
usual Fraunhofer pattern and thus ressembles the interference pattern in a double-slit geometry. 
The distance between two neighboring minima is asymptotically equal to one, the
typical value expected for interference patterns, and thus slightly shorter than the value we have found for the 
small-field regime~$B\ll \Phi_0/d^2$, cf. Eq.~(\ref{minima}).

From Eq.~(\ref{ILargeFlux}), we find 
that the Josephson critical current, Eq.~(\ref{IcDef}),
for~$\varphi$ at around integer~$n$ is given by
\begin{align}
I_c(\varphi) &\simeq 
\frac{8}{9\pi\sqrt{3}}
\frac{I_{c0}}{\varphi}
\frac{\ell_B^2}{d^2}
\big[
1-12\nu^2(\varphi)
\big]^{3/2}\ ,
\label{IMax}
\end{align}
with $\nu(\varphi)=\varphi/2-n$. Formula~(\ref{IMax})
corresponds to ``bell-shaped'' curves as depicted in 
Fig.~\ref{fig_fraunhofer}(c). As Eq.~(\ref{IMax})
describes a maximum of $I(\chi_0,\varphi)$ with respect to~$\chi_0$ for~$\nu(\varphi)$ between $-1/(2\sqrt{3})$ and $1/(2\sqrt{3})$, 
``bells'' from neighboring integer values overlap. Thus, the minima
in~$I_c(\varphi)$ at half-integer values are still kinks, cf. Fig.~\ref{fig_fraunhofer}(c).

The values of the maxima of the critical current, occuring at integer~$\varphi$,
amount to
\begin{align}
\max I_c(\varphi) &\simeq 
\frac{8}{9\pi\sqrt{3}} \frac{I_{c0}}{\varphi}\frac{\ell_B^2}{d^2}
\ ,
\label{IExtreme}
\end{align}
while the minima, situated at half-integer $\varphi$, take the 
values
\begin{align}
\min I_c(\varphi) &\simeq 
\frac{1}{9\pi\sqrt{3}} \frac{I_{c0}}{\varphi}\frac{\ell_B^2}{d^2}
\ .
\label{IExtremeLow}
\end{align}
Minima and maxima are thus related to each other as $\min I_c=\max I_c/8$.
This shows that in the large-field regime, they scale with the magnetic field~$B$ 
in the same manner, as expressed by Eq.~(\ref{Ilargefield}). In particular, $I_c(\varphi)$ features no zeros.

\begin{figure}[t]
\centerline{\includegraphics[width=0.7\linewidth]{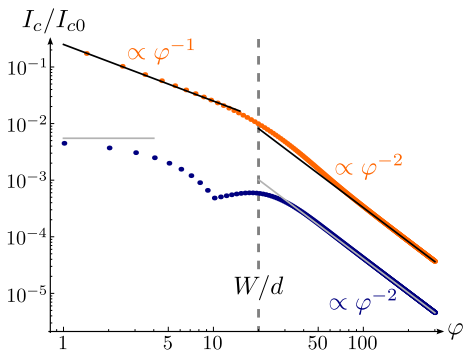}}
\caption{(Color online)
Double-logarithmic representation of maxima and minima of the Josephson critical current as a function of the magnetic flux~$\varphi$.
Dots correspond to values obtained from numerical evaluation of Eqs.~(\ref{I0a}) and~(\ref{I1a}), which are accurate to order $\mathrm{O}(d^2/W^2)$, here for $W/d=20$.
The lines represent the asymptotic behavior as obtained analytically in the low-flux and large-flux regimes, cf. Eqs.~(\ref{maxI_low}), (\ref{minI_low}), (\ref{IExtreme}),
and~(\ref{IExtremeLow}). For magnetic fluxes beyond the crossover at $\varphi\sim W/d$, the critical current falls as $1/\varphi^2$, unlike
the usual Fraunhofer pattern. \emph{Remark}: In the function~$\min I_c(\varphi)$, we observe a kink at~$\varphi\sim 10$. This kink
is explained by the observation
that for~$\varphi\sim W/d$, the current $I(\chi_0,\varphi)$ as a function of~$\chi_0$ develops another extremum upon increasing~$\varphi$ from an integer value. In terms of the illustration of Fig.~\ref{fig_lowflux}(b), this extremum appears at a phase~$\chi_0<\chi_1$ and eventually dominates over the one at~$\chi_2$.
}
\label{fig_largeflux}
\end{figure}

Equations~(\ref{IExtreme}) and~(\ref{IExtremeLow}) complement the asymptotic analysis
of the function~$\eta(\{\varphi\},d^2/\ell_B^2)$, Eq.~(\ref{Ieta}), of the ``modified
Fraunhofer'' pattern describing the Josephson critical current in SNS junctions of 
finite lengths~$d$. Figure~\ref{fig_largeflux} illustrates the two regimes of small and large
magnetic field as well as the crossover at magnetic fields~$B\sim \Phi_0/d^2$.

\section{Disordered edges and mesoscopic fluctuations}
\label{sec:random}

So far, we have been assuming a perfectly clean and rectangular normal layer in
the SNS junction. While clean normal bulks become increasingly realizable in experiment
by using, e.g., (gated) graphene as the normal metal material, the assumption of clean
edges is considerably more difficult to meet experimentally. In this
section, we are relaxing this assumption and consider reflection from edges
of irregular random shape instead.

\subsection{Rough edges}
\label{ssec:random}

\begin{figure}[t]
\centerline{\includegraphics[width=\linewidth]{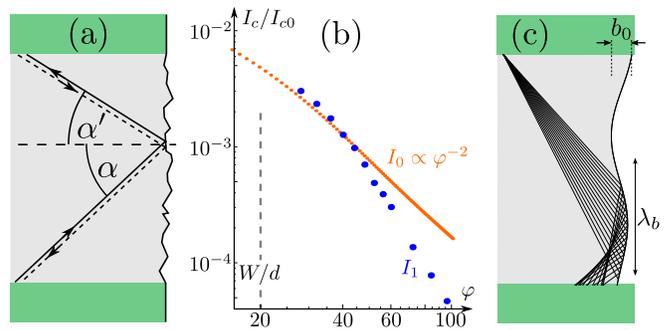}}
\caption{(Color online) (a) Rough side edge model with random reflection angles~$\alpha'$ (measured 
with respect to the perpendicular to a hypothetical clean edge).
(b) Contributions to Josephson critical current maxima due to straight trajectories (thin dots, $I_0$)
    and Fuchs model trajectories (thick dots, $I_1$). At large~$\varphi\gg W/d$ (here $W/d=20$),
    $I_1$ becomes subleading to~$I_0$.
(c) Sketch of a specific realization of edge disorder, for which 
a finite correlation length~$\lambda_b$ leads to a non-linear 
network of semi-classical trajectories. In this situation,
mesoscopic fluctuations can randomly pick currents over length scales
$\gg \ell_B^2/d$, see Sec.~\ref{ssec:classical_meso}, and thus
exceed the universal quantum value~$\delta I_c\sim e/\tau_\mathrm{Th}$, Eq.~(\ref{deltaIc}).
}
\label{fig_rough}
\end{figure}

For very rough side edges, \emph{i.e.}, edge disorder with correlation length much shorter than~$\lambda_F$,
we may follow the early work by Fuchs on the electronic conductivity in thin metallic layers.\cite{fuchs} 
In this model, the roughness of the edge is modeled by treating the reflection angle~$\alpha'$, cf. Fig.~\ref{fig_rough}(a), of each
semi-classical trajectory that collides with the side edges as an independent random variable.
Random trajectories contribute to the current for reflection angles~$\alpha'$ between $0$ and $\pi/2$. We assume
that the random variable~$\alpha'$ is distributed uniformly in this interval and obtain the Josephson current~$I_1(\chi)$,
cf. Eq.~(\ref{Isum}), by averaging over all configurations~$(y,p_y,\alpha')$. 

The current contribution~$I_1(\chi_0,\varphi)$ can be computed using Eq.~(\ref{IGeneral}) with $\tilde{d}$ and~$\delta\chi$
now being functions also of the random variable~$\alpha'$. For positive~$p_y$ and~$y>y^*$, cf. Eq.~(\ref{yc}),
\begin{align}
\tilde{d} &=
\frac{d+(W-2y)\tan\alpha}{2\sin\alpha}+\frac{d-(W-2y)\tan\alpha}{2\sin\alpha'}
\ , \\
\delta\chi &= -\pi\varphi\
    \Big[
    1-\frac{d\cot\alpha-
         (W-2y)}{2W}
    \nonumber\\
    &\qquad\qquad\qquad -
      \frac{d\cot\alpha + (W-2y)}{2W}\frac{\cot\alpha'}{\cot\alpha}
    \Big]
\ ,
\label{random_d_and_chi}
\end{align}
which for diffusive reflection with $\alpha'\neq\alpha$ differs from Eqs.~(\ref{deff}) and~(\ref{phasedifference_1}).
Averaging over $\alpha'$ has to be done for each phase space point
$(y,p_y)$. Then, the additional integration over~$\alpha'$, as we may expect,
should lead to an extra factor of order~$(W/d)/\varphi\ll 1$ in the regime of large flux~$\varphi\gg W/d$.
This is in fact confirmed by numerical simulations, cf. Fig.~\ref{fig_rough}(b).
Asymptotically, for $\varphi\gg W/d$, we thus find that the contribution due to trajectories
including reflection from rough edges is smaller than the contribution due to straight trajectories,
$I_1\sim I_0 (W/d)/\varphi \ll I_0$. As a result, only
the latter contribution persists. This leads to
\begin{align}
\max I_c(\varphi) &\simeq 
\frac{4}{9\pi\sqrt{3}} \frac{I_{c0}}{\varphi}\frac{\ell_B^2}{d^2}
\ ,\,\,\,
\min I_c(\varphi) \simeq 
\frac{1}{8}\max I_c(\varphi)
\label{Irandom}
\end{align}
instead of the clean-edge results~(\ref{IExtreme}) and~(\ref{IExtremeLow}), to which
$I_1$ contributed equally. We find that due to roughness in the edges, the (average) Josephson current is smaller
by a factor of (at most) two. Our earlier results building on the assumption of clean edges 
thus remain qualitatively valid. 

For small flux, $\varphi\ll W/d$, Eqs.~(\ref{maxI_low}) and~(\ref{minI_low}) are found to
remail valid apart from minor alterations of the numerical coefficient~$f_0$.

\subsection{Mesoscopic fluctuations}

Until now, our study has addressed the average Josephson critical current~$I_c(\varphi)$.
Disorder, as induced by the randomness in the edges, results in sample-to-sample (mesoscopic)
fluctuations\cite{akkermans} that blur the average value~$I_c$. 
If the mesoscopic fluctuations become strong, corresponding to a standard deviation~$\delta I_c\gtrsim I_c$,
the average value is no longer a representative experimental quantity.

According to Eqs.~(\ref{IExtreme}), (\ref{IExtremeLow}), and~(\ref{Irandom}), the average current varies with flux as~$I_c\propto 1/\varphi^2$. In contrast, as we will see, $\delta I_c$ drops more slowly with increasing~$\varphi$. 
This results in the existence of a crossover scale~$\varphi_\sigma$ in the magnetic flux,
at which~$I_c(\varphi_\sigma)\sim \delta I_c(\varphi_\sigma)$. For larger flux, mesoscopic fluctuations
dominate the Josephson current.
In this section, we are providing estimates for the scale~$\varphi_\sigma$.

\subsubsection{Quantum mesoscopic fluctuations}

Randomness in the edge can be characterized by means of a correlation length~$\lambda_b$.
In the Fuchs model used in Sec.~\ref{ssec:random}, reflection angles of any pair of different trajectories 
are uncorrelated, which corresponds to a vanishing correlation length~$\lambda_b=0$. In 
this situation, mesoscopic fluctuations vanish as well. This remains essentially true as long as~$\lambda_b\ll\lambda_F$.

For edge disorder with correlation length~$\lambda_b\sim\lambda_F$,
quantum effects start inflicting fluctuations~$\delta I_c$ to the critical Josephson current.
As these arise in narrow strips close to the edges at $y=\pm W/2$,
we may estimate their order of magnitude by adopting the known results
from one-dimensional SNS junctions,\cite{altshulerspivak,beenakker91,houzet}
\begin{align}
\delta I_c \sim \frac{e}{\tau_{\mathrm{Th}}}
\ ,
\label{deltaIc}
\end{align}
where $\tau_{\mathrm{Th}}$ is the electron traversal ({\it i.e.}, Thouless) time.
We use $\tau_{\mathrm{Th}}\sim d/v_F$, as we assume
ballistic transport through the normal-state region.

The fluctuations~$\delta I_c$, Eq.~(\ref{deltaIc}), are notably independent of the magnetic field~$B$.
Therefore, at large enough fields, they will eventually dominate over the average value~$I_c$, which drops as $1/B^2$.
The average value~$I_c$ remains meaningful as long as~$B\ll B_\sigma^{\rm qu}$, with the crossover scale given by
\begin{align}
B_\sigma^{\rm qu}
 \sim
 \frac{\Phi_0}{d^2} \sqrt{\frac{d}{\lambda_F}}
\label{phic} \ .
\end{align}
Under the conditions~(\ref{limits}), the crossover to strong mesoscopic fluctuations occurs deep in the regime of the dominant edge effect, $B_\sigma^{\rm qu}\gg\Phi_0/d^2$. 

Expressing the magnetic field~$B_\sigma^{\rm qu}$, Eq.~(\ref{phic}), in terms
of the number~$n_\sigma$ of filled Landau levels, we find $n_\sigma\sim(d/\lambda_F)^{3/2}\gg 1$. Thus, mesoscopic fluctuations
become dominant for magnetic fields much lower than those needed to enter a quantum Hall regime of (Andreev) edge transport. It means
that the quantum Hall effect in an SNS junction develops in a non-universal way determined by mesoscopic fluctuations.

\subsubsection{Classical mesoscopic fluctuations}
\label{ssec:classical_meso}

If the shape of the disordered edge varies on a characteristic scale~$\lambda_b$ exceeding the ``quantum''
scale $\sqrt{d\lambda_F}$, the crossover at~$B_\sigma^{\rm qu}$ specified by Eq.~(\ref{phic}) 
is replaced by a crossover into a regime of \emph{classical mesoscopic fluctuations}
occurring at a weaker field~$B_\sigma^{\rm cl}$. In the picture of semi-classical trajectories, which is analogous to geometric optics, 
a edge smoothly varying on a scale~$\lambda_b$ works as a curved mirror, cf. Fig.~\ref{fig_rough}(c). 
Specular reflection from it, in general, introduces focal points and caustics into the ray optics, as illustrated in the figure:
an ensemble of rays originating from the same point at the far SN interface does not strike the other interface homogeneously
but tends to bunch into inhomogeneously distributed ``speckles''. The random arrangement of speckles
is a possible source of enhanced mesoscopic fluctuations in the current.

In order to study the classical mesoscopic fluctuations analytically, we choose the following
model of a specular fluctuating edge with random curvature: 
introducing a function~$b(x)$ with values of order~$b_0\ll d$ and varying on the scale~$\lambda_b\ll d$, cf. Fig.~\ref{fig_rough}(c), 
we describe the edge at~$y=W/2$ to be the graph of the function~$x\mapsto y=W/2+b(x)$. The edge function~$b(x)$
is assumed to be random with a Gaussian distribution characterized by zero mean, $\langle b(x)\rangle =0$, and
auto-correlation
\begin{align}
\left\langle
b(x)b(x')
\right\rangle
= b_0^2 \mathrm{e}^{-|x-x'|/\lambda_b}
\ .\label{bcorrelation}
\end{align}
We furthermore assume $b_0\ll\lambda_b$, which implies that the local angle~$\beta(x)$ between the disordered edge
and the straight normal line is small,
\begin{align}
\beta(x)\simeq\partial_x b(x)\ll 1
\ .
\label{beta}
\end{align}
For a given configuration~$\{b(x)\}$ of the edge, the Josephson current~$I_1(\chi,\varphi)$ due to trajectories 
with side edge scattering is then obtained from Eq.~(\ref{IGeneral}) with the spatial integral restricted
to $y>y^*$, cf. Eq.~(\ref{yc}), and where $\tilde{d}$ and~$\delta\chi$ are given by Eq.~(\ref{random_d_and_chi}).
Here, however, the angle~$\alpha'$ entering the effective length of the trajectory and phase difference is 
explicitly given by 
\begin{align}
\alpha' &= \alpha + 2\beta\big((\tfrac{1}{2}W-y)\tan\alpha\big)
\ ,
\end{align}
corresponding to specular reflection with respect to the local orientation of the edge function~$b(x)$.
Disorder averaging is carried out by averaging over all configurations~$\{b(x)\}$ using Eq.~(\ref{bcorrelation}).

The analytical procedure specified above in principle allows us to calculate the average Josephson current~$\langle I_c\rangle$ as
well as any moment~$\langle I_c^n\rangle$. Practically, the non-analyticity of the function~$J(\chi)$, Eq.~(\ref{IZeroB}),
constitutes a complication, which we may heal by choosing the analytic ``model function'' $J(\chi)=I_{c0}\sin\chi$ instead.
Neglecting all harmonics higher than the first one is an uncontrolled approximation but should nevertheless provide
the correct order of magnitude for estimates.\cite{footnote01} In particular, if we had adopted this approximation throughout 
our analysis, we would have encountered the same qualitative crossover behavior of the average current at~$B\sim\Phi_0/d^2$ as found
in the preceding sections. The assumed analytic shape
of~$J(\chi)$ simplifies the analysis as it allows for simple expansions in small quantities such as~$\beta(x)$, Eq.~(\ref{beta}).

Fluctuations have most drastic consequences close to the minima of the critical current. Setting~$\chi=\pi/2$ and assuming 
half-integer~$\varphi$ (in the regime $\varphi\gg W/d$), we find using the above described procedure that the classical
mesoscopic fluctuations are characterized by standard deviation
\begin{align}
\delta I_c \simeq \frac{4\sqrt{2}}{\pi^2}\frac{I_{c0}}{\varphi}\frac{b_0}{d}
\ ,\label{deltaIc_cl}
\end{align}
cf. Eq.~(\ref{deltaIcintro}).
Note that $\delta I_c\propto 1/B$ falls off with the magnetic field slower than the average current does, cf. Eqs.~(\ref{IExtreme}) and~(\ref{IExtremeLow}).
Whereas Eq.~(\ref{deltaIc_cl}) formally is independent of the disorder correlation length~$\lambda_b$, 
an implicit dependence is given by the fact that it has been derived assuming~$b_0\ll\lambda_b$. At the border of applicability, the 
typical angle~$\beta_0\sim b_0/\lambda_b\sim 1$, and Eq.~(\ref{deltaIc_cl}) predicts a~$\delta I_c\propto \lambda_b$ scaling.

The mesoscopic fluctuations~(\ref{deltaIc_cl}) become of the same order as the maxima of the critical current~$I_c(\varphi)$, cf. Eq.~(\ref{IExtreme}),
for magnetic fields~$B\sim B_\sigma^{\rm cl}$ with
\begin{align}
B_\sigma^{\rm cl}\sim \frac{\Phi_0}{d^2}\frac{d}{b_0}\sim \frac{\Phi_0}{d^2}\frac{d}{\beta_0\lambda_b}
\ .\label{phic_cl}
\end{align}
Comparing it with Eq.~(\ref{phic}), we see that indeed $B_\sigma^{\rm cl}\lesssim B_\sigma^{\rm qu}$ if the parameter~$b_0=\beta_0\lambda_b$ is greater than~$\sqrt{\lambda_Fd}$. In either case, under the conditions~(\ref{limits}), the crossover to strong mesoscopic fluctuations occurs within the regime of edge-dominated transport, 
\emph{i.e.}, $\min(B_\sigma^{\rm cl},B_\sigma^{\rm qu})\gg \Phi_0/d^2$.

\section{Discussion}
\label{sec:discussion}

In this work, we have studied the Josephson current through a long and wide SNS junction
 with large but finite aspect ratio~$W/d$ that is penetrated by an external magnetic field. 
Assuming clean and perfectly rectangular junctions, we have identified a crossover between two regimes of
the Josephson critical current~$I_c$ as a function of the magnetic
field, which takes place as the magnetic length~$\ell_B$ and the junction width~$d$ become of
comparable size, $\ell_B\sim d$ or, in other words $B\sim \Phi_0/d^2$. 
The results of our analysis of the Josephson critical current in the two regimes are summarized in Eq.~(\ref{Ieta}),
which notably modifies the functional dependences known from Fraunhofer patterns. The ``modified Fraunhofer''
pattern in the two regimes and the crossover region is graphically illustrated in Figs.~\ref{fig_fraunhofer} and~\ref{fig_largeflux},
obtained by a numerical simulation.

In the crossover region, $B\sim\Phi_0/d^2$, besides the gradual change of the power law $\max I_c\propto 1/B^\gamma$ from $\gamma=1$
to $\gamma=2$, the ``period'' of the Fraunhofer pattern, \emph{i.e.}, the distance between neighboring minima of $I_c(\Phi/\Phi_0)$,
passes from a geometry-dependent value slightly larger than one, cf. Eq.~(\ref{minima}), to the universal value of one.
In fact, our numerics indicate that the distance between critical current minima changes non-monotonously as a function of~$B$: whereas
for $B\ll\Phi_0/d^2$ it first slightly grows, it starts shrinking down to one only at $B\sim\Phi_0/d^2$ 
(specifically, beyond the kink of the minimal current shown in Fig.~\ref{fig_largeflux}). We note that 
in a recent study\cite{rakyta} a similar pattern has been observed in simulations of Josephson transport
in moderately wide junctions. 

The change of the power law and the variation of period in the interference~$B\sim\Phi_0/d^2$ constitute verifiable experimental signatures 
of our theory. In order to fit experimental results, the Josephson critical current as
obtained by using Eqs.~(\ref{I0a}) and~(\ref{I1a}) may be advantageous over the standard
Fraunhofer formula\cite{tinkham} also for magnetic fields~$B\lesssim\Phi_0/d^2$.

We have shown that disorder in the side edges
may, to leading order, affect the numerical coefficients in
the expression for the (average) Josephson critical current by reducing it at most by a factor of two, cf. Eq.~(\ref{Irandom}).
This means, however, that it does not qualitatively change the asymptotic results that had been obtained for clean edges.
In particular, there is no exponential decay as predicted for diffusive SNS junctions\cite{ivanov} in the regime~$B\gtrsim\Phi_0/d^2$.

Mesoscopic fluctuations originating from such disordered scattering, however,
limit the validity of the theory to magnetic fields below a certain value~$B_\sigma$, which 
for typical parameters, cf. Eq.~(\ref{limits}),
is much larger than the crossover magnetic field~$\sim\Phi_0/d^2$.
We distinguish mesoscopic fluctuations of universal quantum origin, cf. Eqs.~(\ref{deltaIc}) and~(\ref{phic}),
and classical mesoscopic fluctuations due to specular reflection from a randomly curved edge varying
on a characteristic length scale~$\lambda_b>\sqrt{\lambda_Fd}$, cf. Eqs.~(\ref{deltaIc_cl}) and~(\ref{phic_cl})
and Fig.~\ref{fig_rough}(c).
The presence of mesoscopic fluctuations particularly implies a non-universal crossover into the quantum Hall regime
at very large magnetic fields.

Another limit of applicability of our results, Eqs.~(\ref{result_max}) and~(\ref{result_min}), originates from
the initial assumption of straight semi-classical trajectories in the normal layer.
In fact, the curvature of cyclotron orbits at non-zero magnetic fields prevents
the Andreev-reflected part of a trajectory from retracing the incident trajectory.
This seems to constitute a major obstacle to the formation of Andreev bound states along
one-dimensional semi-classical trajectories. 
At the same time, we note that the semi-classical treatment built furthermore
on the assumption of coordinate~$y$ and momentum~$p_y$ along
the SN interface being classical variables. Quantum uncertainty in
these quantities can restore the Andreev bound states if
the magnetic field~$B$ is not too large, $B\lesssim B^*$.\cite{falkogeim}

\begin{figure}[t]
\centerline{\includegraphics[width=0.45\linewidth]{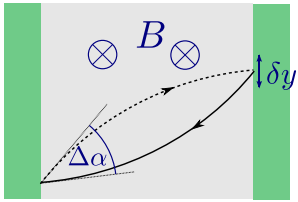}}
\caption{(Color online) 
Semi-classical orbits at non-zero magnetic field. The finite curvature 
of the cyclotron trajectories leads to an angular mismatch~$\Delta\alpha$ at
the SN interfaces. At small enough magnetic fields, $B\lesssim B^*$,
this mismatch is unimportant compared to quantum uncertainty but impedes
the formation of Andreev states at larger fields.
}
\label{fig_curvature}
\end{figure}

In order to estimate the scale~$B^*$, consider 
an electron-hole pair of classical cyclotron orbits
connecting the same points along the SN interfaces, see Fig.~\ref{fig_curvature}.
Whereas the formation of (low-lying) Andreev states between these points requires
that angles of incidence and reflection be the same,
the curvature of the cyclotron orbits enforces an angular mismatch~$\Delta\alpha$.
Assuming not far from normal incidence, we estimate that typically 
$\Delta\alpha\sim d/r_B$, where $r_B$ is the cyclotron radius. 
On the other hand, quantum uncertainty in~$y$ and~$p_y$ induces uncertainty
also in the angle between incident and reflected trajectories. Noting that
spatial uncertainty~$\delta y$ is typically of the order of the superconducting coherence length, 
\emph{i.e.}, $\delta y\sim\xi$, we find an angular quantum uncertainty of~$\delta\alpha\sim\lambda_F/\xi$.
As long as the angular mismatch~$\Delta\alpha$ due to cyclotron orbits
remains smaller than~$\delta\alpha$, we may consider it unimportant from a quantum point of view.
Thus the upper bound for magnetic fields allowing for the
formation of Andreev states is found\cite{falkogeim} as
\begin{align}
B^*\sim\frac{\Phi_0}{d^2}
     \frac{d}{\xi}
\label{Bstar}\ .
\end{align}
Note that in the limits~(\ref{limits}) assumed
in our theory, $B^*$ is larger than magnetic fields at the crossover into
the regime of edge-dominated Josephson transport, $B^*\gg\Phi_0/d^2$.

Depending on whether $B^*<B_\sigma$ or $B^*>B_\sigma$,
see Eqs.~(\ref{phic}) and~(\ref{phic_cl}), either the curvature of electronic orbits in the magnetic field 
or mesoscopic fluctuations constitute the primary limit of the applicability of our results
for the average Josephson critical current.

\section{Conclusion}
\label{sec:conclusion}

We have studied the Josephson critical current~$I_c$ in a two-dimensional long and wide SNS junction
threaded by magnetic flux~$\Phi$, assuming a clean bulk material but allowing for disordered edges. 
Our results indicate that scattering off the side edges induces an additional parametric dependence 
of~$I_c$ on the ratio $\ell_B/d$ of the magnetic length to the junction width, which is absent in the usual Fraunhofer
pattern in SIS or short SNS junctions. The crossover at~$\ell_B/d\sim 1$, which separates the two
regimes studied in the present work, should lead to clear signatures within reachable experimental scales. We hope
that the ``modified Fraunhofer'' pattern predicted by our theory will be useful for fitting future experimental data.

\acknowledgments

We acknowledge useful discussions with Mosche Ben Shalom, Richard T. Brierley, Andre Geim, Manuel Houzet, Angela Kou, Ivana Petkovi\'{c}, and Mengjian Zhu.
The work at Yale was supported by ARO Grant W911NF-14-1-0011 and 
NSF DMR Grants No. 1206612 (L.~G.) and 1301798 (H.~M.).
V.~F. was supported by the European Graphene Flagship Project and the ERC Synergy Grant Hetero2D.
H.~M. acknowledges the Yale Prize Postdoctoral
Fellowship. H.~M. and L.~G. thank the University of Manchester Visiting Program in Theoretical Physics for hospitality.


\begin{thebibliography}{99}
\bibitem{footnote01}
We emphasize that this is not a high-temperature approximation.
At large temperatures, $k_BT\gg v_F/d$, as is well-known, the Josephson current~(\ref{IZeroB})
through long junctions effectively reduces to sinusoidal shape. The critical current, however,
acquires an additional $\tilde{d}$-dependence in form of an exponential factor~$\exp[-2\pi k_BT/(v_F/\tilde{d})]$.
\end{thebibliography}
\end{document}